\documentclass{osa-article}
\usepackage{physics}
\journal{oe}

\articletype{Research Article}

\begin{document}

\title{Performance of real-time adaptive optics compensation in a turbulent channel with high-dimensional spatial-mode encoding}

\author{Jiapeng Zhao,\authormark{1,*} Yiyu Zhou,\authormark{1} Boris Braverman, \authormark{2} Cong Liu, \authormark{3} Kai Pang, \authormark{3} Nicholas K. Steinhoff, \authormark{4} Glenn A. Tyler, \authormark{4} Alan E. Willner \authormark{3} and Robert W. Boyd \authormark{1,2}}

\address{\authormark{1}The Institute of Optics, University of Rochester, Rochester, New York, 14627, USA\\
\authormark{2}Department of Physics, University of Ottawa, Ottawa, Ontario, K1N 6N5, Canada\\
\authormark{3}Department of Electrical and Computer Engineering, University of Southern California, California, 90007, USA\\
\authormark{4}The Optical Science Company, Anaheim, California, 92806, USA}

\email{\authormark{*}jzhao24@ur.rochester.edu} 



\begin{abstract}
 The orbital angular momentum (OAM) of photons is a promising degree of freedom for high-dimensional quantum key distribution (QKD). However, effectively mitigating the adverse effects of atmospheric turbulence is a persistent challenge in OAM QKD systems operating over free-space communication channels. In contrast to previous works focusing on correcting static simulated turbulence, we investigate the performance of OAM QKD in real atmospheric turbulence with real-time adaptive optics (AO) correction. We show that, even our AO system provides a limited correction, it is possible to mitigate the errors induced by weak turbulence and establish a secure channel. The crosstalk induced by turbulence and the performance of AO systems are investigated in two configurations: a lab-scale link with controllable turbulence, and a 340 m long cross-campus link with dynamic atmospheric turbulence. Our experimental results suggest that an advanced AO system with fine beam tracking, reliable beam stabilization, precise wavefront sensing, and accurate wavefront correction is necessary to adequately correct turbulence-induced error. We also propose and demonstrate different solutions to improve the performance of OAM QKD with turbulence, which could enable the possibility of OAM encoding in strong turbulence.
\end{abstract}

\section{Introduction}

\paragraph{}Quantum key distribution (QKD), which assures unconditionally secure communication between multiple parties, is one of the most promising and encouraging applications of quantum physics \cite{gisin2002quantum,scarani2009security,lo2014secure}. Instead of relying on mathematical complexity, the security of QKD is guaranteed by fundamental physical laws, which indicate that the encrypted keys will remain secure even against eavesdroppers with unlimited computation power \cite{gisin2002quantum,scarani2009security,lo2014secure}. \par

Since its birth in 1984 \cite{bennett1984quantum}, the concepts of QKD have been demonstrated in various platforms, including fiber-based networks \cite{korzh2015provably, yin2016measurement}, free-space communication links \cite{liao2017satellite, jin2019genuine}, underwater \cite{ji2017towards, bouchard2018quantum} and over-marine channels \cite{ursin2007entanglement, sun2016multiple}. However, in most QKD systems, the information is encoded in the polarization degree of freedom, which is a two-dimensional Hilbert space limiting the information capacity to 1 bit per photon. Even through a single-photon source with a high brightness has been developed \cite{dousse2010ultrabright,steinlechner2012high}, the two-dimensional QKD systems are still photon-inefficient. 

As a comparison, high-dimensional QKD systems are more photon-efficient and robust to eavesdropping \cite{cerf2002security,sheridan2010security,bradler2016finite}. In recent decades, many new protocols involving high-dimensional encoding have emerged. Encoding information with orbital angular momentum (OAM) states, which can span an infinite-dimensional Hilbert space, has been experimentally demonstrated to be advantageous in both high-dimensional quantum cryptography \cite{vallone2014free, mirhosseini2015high, sit2017high, bouchard2018experimental, bouchard2018round} and classical communication \cite{wang2012terabit, zhu2019compensation}. By definition, an OAM state $\ket{\ell}$ carrying $\ell\hbar$ units of OAM has $\ell$ intertwined helical wavefronts, where $\ell$ denotes the OAM quantum number and is an integer \cite{allen1992orbital}. While efficient and high-fidelity fibers for high-order spatial modes are under investigation \cite{flaes2018robustness, cozzolino2019orbital}, OAM QKD in free-space links remains desirable due to the greater flexibility in applications and the lower loss. Since the information is carried by the phase profile, OAM states are vulnerable to atmospheric turbulence. Even though the performance of OAM states in a turbulent channel has been studied both theoretically and experimentally \cite{tyler2009influence, tyler2011spatial, ren2013atmospheric, ren2014adaptive, rodenburg2014simulating, ren2016experimental, bouchard2018quantum, liu2019single, zhao2019performance, zhao2019experimental, hufnagel2019characterization, li2019mitigation}, realizing high-dimensional OAM-based QKD still remains challenging. \par 

To reduce the crosstalk induced by turbulence, most works either rely on post-selection of data or increasing the mode spacing (i.e. not using successive states for encoding) \cite{ren2016experimental, sit2017high, bouchard2018quantum}. For a given free-space link, although these methods can reduce the quantum symbol error rates (QSER), they lead to a reduction of photon rate and size of encoding space. Therefore, the advantage of high-dimensional encoding on information capacity cannot be fully realized. Moreover, for an OAM encoding space with mode spacing equal to one, the OAM basis and angular (ANG) basis form the mutually unbiased bases (MUBs). The ANG basis can be described as:
\begin{equation}
\ket{j} = \frac{1}{\sqrt{d}}\sum_{\ell = -L}^{L}\ket{\ell}\text{exp}(-i2\pi j\ell/(2L+1)),
\end{equation}
where $L$ is the maximal OAM quantum number in use. A high-fidelity sorter for efficiently measuring these MUBs has been developed, and its effectiveness has been demonstrated in QKD systems as well \cite{mirhosseini2013efficient, mirhosseini2015high, fu2018realization}. However, its counterparts for mode spacing larger than one have not yet been demonstrated, and an inefficicent measurement device may introduce additional security loopholes \cite{acin2016necessary}. Therefore, an efficient approach which can both take the full advantages of high-dimensional encoding and reduce the crosstalk from atmospheric turbulence is still under investigation. \par

Conceptually, the simplest technique for overcoming turbulence-induced errors is to use adaptive optics (AO) to correct the phase errors and recover the benefits of using the high-dimensional encoding QKD system. However, since OAM states are very sensitive to wavefront errors, any imperfect correction may actually lead to an increase rather than a decrease in QSER, and hence the failure of QKD system. Most relevant experimental works only focus on correcting static turbulence simulated by single or multiple random phase screens, which is a significantly simplified model ignoring the dynamic nature of atmospheric turbulence \cite{ ren2013atmospheric,rodenburg2014simulating, ren2014adaptive, liu2019single}. In addition, some theoretical simulations predict that a simple AO system may not be adequate for turbulence correction \cite{GlennSlides, GlennBook, lavery2018vortex}. Therefore, under real dynamic turbulence, using AO systems to correct errors in OAM states remains very challenging, and the performance of AO correction in real atmospheric turbulence is still unknown and needs to be investigated. \par

To thoroughly study the effect of AO correction on OAM states, we investigate the performance of an OAM QKD system with real-time AO compensation in both a lab-scale and a cross-campus link. We first quantitatively investigate the performance of such a QKD system in the lab with a controllable source of turbulence. We find out that, even though AO only provides a limited correction, the quantum channel disturbed by weak turbulence can remain secure when the compensation is enabled. We then study the performance of OAM QKD in a 340 m long cross-campus link. Due to the relatively high turbulence level and modest performance of AO, we can reduce the QSER in the cross-campus link but it is still too high to guarantee the security of the channel. Based on our observation and previous simulation results, advanced AO system with fine beam tracking, reliable beam stabilization, precise wavefront sensing and accurate wavefront correction is necessary to correct the error induced by moderate or strong turbulence. In our summary, we propose three different solutions to improve the performance, and show that the performance of spatial mode QKD system can be improved if these methods are implemented.

\section{Method and Results}
\subsection{Lab-scale link under controllable turbulence}

\begin{figure*}[ht]

\centering
\fbox{\includegraphics[width=0.95\linewidth]{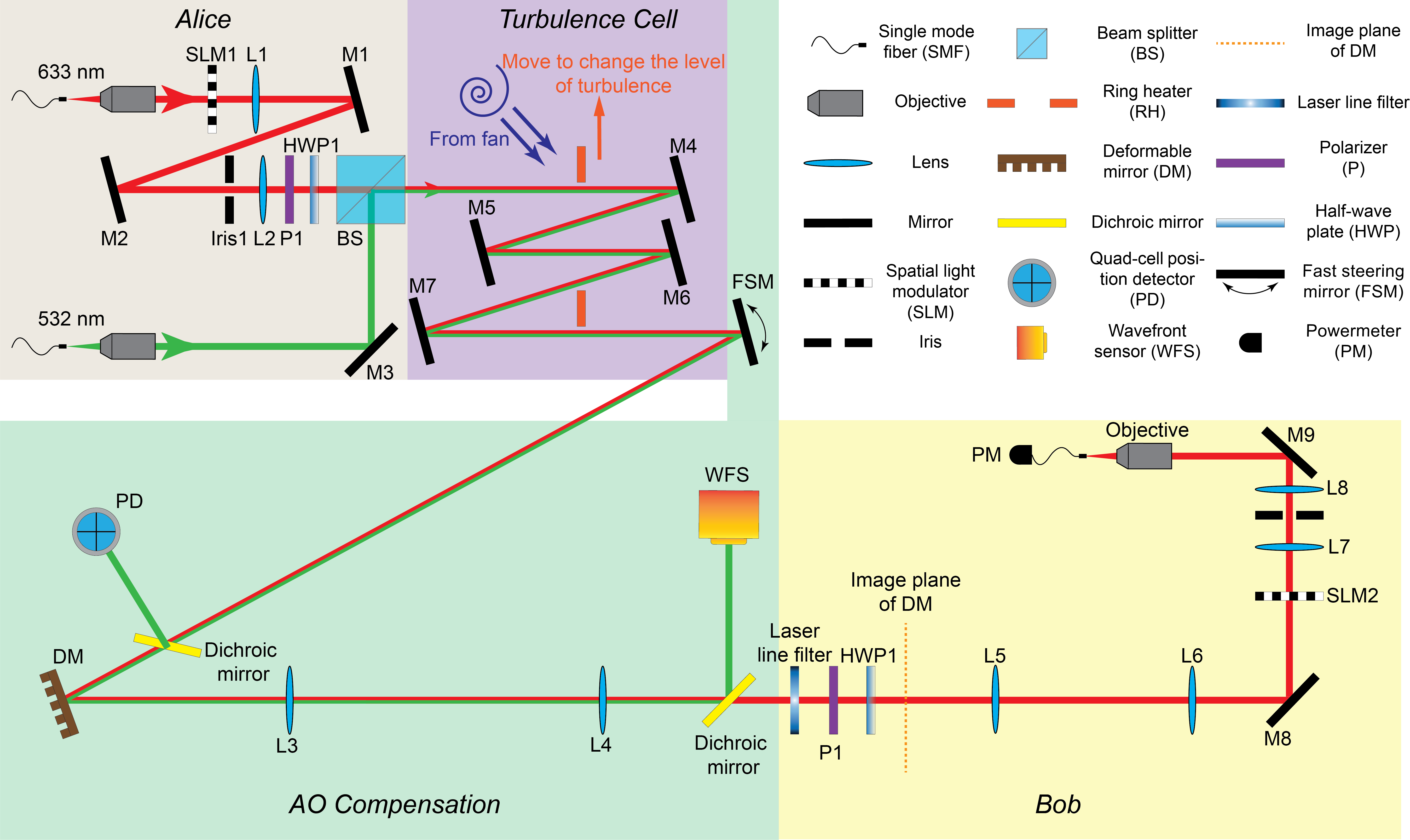}}
\caption{The configuration of the lab-scale link with a controllable turbulence cell. Both signal and beacon beams go through the center of the RH. The size of the signal beam is selected to cover the central part of the DM (3$\times$3 actuators) to avoid the cutoff from the edges. The size of the beacon beam overfills the DM aperture to provide a precise estimation of turbulence across the DM. The polarization of the signal beam is controlled by a polarizer to encode information while the polarization of the beacon beam is fixed in $\ket{H}$ state.}
\label{fig:fig1}

\end{figure*}

We first investigate the influence of atmospheric turbulence on OAM states and the conjugate ANG states under different levels of turbulence, and then perform real-time AO compensation on these states. The experimental setup is shown in Fig. 1. Alice prepares her states using a spatial light modulator (SLM) and a 633 nm He-Ne laser. The laser is first coupled into a single mode fiber (SMF) to generate a fundamental Gaussian state ($\ell = 0$), which is then collimated with an objective and illuminates SLM1 (SDE1024 from Cambridge Correlators Ltd). A pair of lenses ($f_1$ = 0.75 m and $f_2$ = 0.5 m) together with an iris are used to select the desired state of light, carried by the first-order diffraction from the SLM. A polarizer and a half-wave plate (HWP1) after these lenses are used to prepare four different polarization states: horizontal ($\ket{H}$), vertical ($\ket{V}$), diagonal ($\ket{D}$) and anti-diagonal ($\ket{A}$). The beacon beam, which comes from a 532 nm green laser, is collimated using an aspheric lens from a SMF. The signal beam (the beam encoded by SLM1) is combined with the beacon beam through the use of a beam splitter (BS). Afterwards, both beams propagate collinearly through the turbulent channel consisting of a turbulence cell (TC) and three mirrors. The TC is a ring heater (RH) blown on by a fan. We adjust the level of turbulence by changing three parameters: the temperature of RH, the fan speed and the number of times that the beams go through the TC. The separation between BS and RH is 1.5 m while the separations between RH and the first two mirrors (M4 and M5) are both 0.15 m. For the strongest turbulence, the beams go through the TC four times and are then reflected to the deformable mirror (DM) by a fast steering mirror (FSM, OIM5002 from Optics In Motion LLC). For the weakest turbulence, RH is moved 0.3 m away from the beams so that the beams simply bypass the RH but still experience some turbulence from the edge of RH.\par

The AO compensation system consists of two parts. The first part has a FSM and a quad-cell position detector (PD), and is used to correct the beam wander induced by the 2nd and 3rd Zernike polynomials (in Noll index, i.e. tip and tilt). To redirect part of the green beacon beam to PD, a 488 nm 50/50 non-polarizing BS (\#48-217 from Edmund Optics) is used as a dichroic mirror, which leads to 7.40$\%$ loss in the signal beam. Since one set of FSM and PD is involved, only two degrees of freedom can be corrected (either x-y position or the propagation direction on the DM, i.e. the x-y momentum). In our configuration, the beam position on the DM is corrected but not the propagation direction. To minimize the tip-tilt error on DM induced by the FSM, the separation between FSM and PD is much larger than the separation between FSM and TC. That is to say, the FSM is in the near-field of turbulence while the PD is in the far field. The second part of the AO system consists of a Shack-Hartmann wavefront sensor (WFS, WFS20-7AR from Thorlabs) and a DM with 32 actuators (DM32-35-UM01 from Boston Micromachine). WFS is working in the high-speed mode with 23$\times$23 microlens in use, and the measured Zernike coefficients are limited to the first 15 terms to give the best performance. The selection of this optimal specification will be discussed later. The compensation control is performed by Thorlabs AO kit software (Version 4.40). To get the optimal wavefront measurement, the beams at DM plane are imaged onto WFS plane using two Thorlabs best-form spherical singlet lenses ($f_3$ = 0.20 m and $f_4$ = 0.15 m). Before the WFS, a 605 nm dichroic mirror (\#34-740 from Edmund Optics) is used to reflect the green beacon beam but transmit the red signal beam. To reduce the noise from beacon beam \cite{ren2017spatially}, a laser line filter (\#68-943 from Edmund Optics) at 633 nm is used to filter out the residual green light. The DM plane is then imaged onto SLM2 using another imaging system, consisting of two Thorlabs best form spherical singlet lenses ($f_5$ = 0.2 m and $f_6$ = 0.2 m), to perform projective measurements \cite{zhao2019performance}. To measure the polarization degree of freedom, a polarizer and a half-wave plate (HWP2) are used after the laser line filter. HWP2 is used to rotate the polarization state ($\ket{H}$, $\ket{V}$, $\ket{D}$ or $\ket{A}$, which can be used as another degree of freedom in hybrid encoding and will be discussed later) to $\ket{H}$ since liquid crystal SLM only affects horizontally polarized light. \par

 To quantify the level of turbulence, we first introduce the Fried parameter $r_0$ which is the spatial coherence length of atmospheric eddies (called turbules), and then use the quantity $D/r_0$ to describe the level of turbulence, where $D$ is the beam diameter because the beacon Gaussian beam underfills the collection aperture \cite{fried1965statistics, fried1966optical}.  Therefore, the quantity $D/r_0$ denotes the number of turbules inside the beam cross section. A large $D/r_0$ indicates strong turbulence ($D/r_0>1$) while a small $D/r_0$ stands for weak turbulence ($D/r_0<1$) \cite{andrews2005laser}. In our experimental setup, the OAM states with quantum number $\ell$ from $-2$ to 2 comprise our encoding space with dimension $d = 5$. The $r_0$ under different levels of turbulence are estimated from the beam wander at the receiver side, which leads to an experimental $D/r_0$ from 0.11 to 3.06. This range spans turbulence levels from weak turbulence to strong turbulence. \par

We measure the crosstalk matrices between the prepared and received states under different turbulence situations, and calculate the measured fidelity ($F$) as a function of turbulence strength $D/r_0$, which are shown in Fig. 2. With the turbulence turned off, we measure an average fidelity $F$ = 93.69$\%$ of the MUBs (the blue dot in Fig. 2(c)). The fidelity of the OAM basis and ANG basis can be found in Fig. 2(a) and (b) respectively. Each measurement takes 1.5 mins so that the total measurement time for the average fidelity of the MUBs under one specific level of turbulence is about 150 mins. The measured fidelity of the MUBs is above the fidelity threshold $F$ = 79.01$\%$ for this $d = 5$ system, which indicates that a secure quantum channel can be established \cite{cerf2002security, sheridan2010security}. As $D/r_0$ increases, $F$ drops quickly due to an increase in fluctuation levels, and the fidelity in the OAM basis matches well with the theoretical prediction $F = 1- [1+c(D/r_0)^2]^{-1/2}$, where coefficient $c$ is 3.404 for the no turbulence case \cite{tyler2009influence}. The yellow and red curves are least square fitting results of the experimental data. Even under weak turbulence with $D/r_0$ = 0.11, the average fidelity in the ANG basis (78.04$\%$) is below the threshold. After we turn on the AO, the fidelity is improved to 80.07$\%$ in the ANG basis and the fidelity in the OAM basis is improved from 86.69$\%$ to 90.35$\%$. Therefore, a secure channel, which could not have been otherwise established, becomes possible after the AO correction is applied. \par

\begin{figure}[ht]

\centering
\fbox{\includegraphics[width=\linewidth]{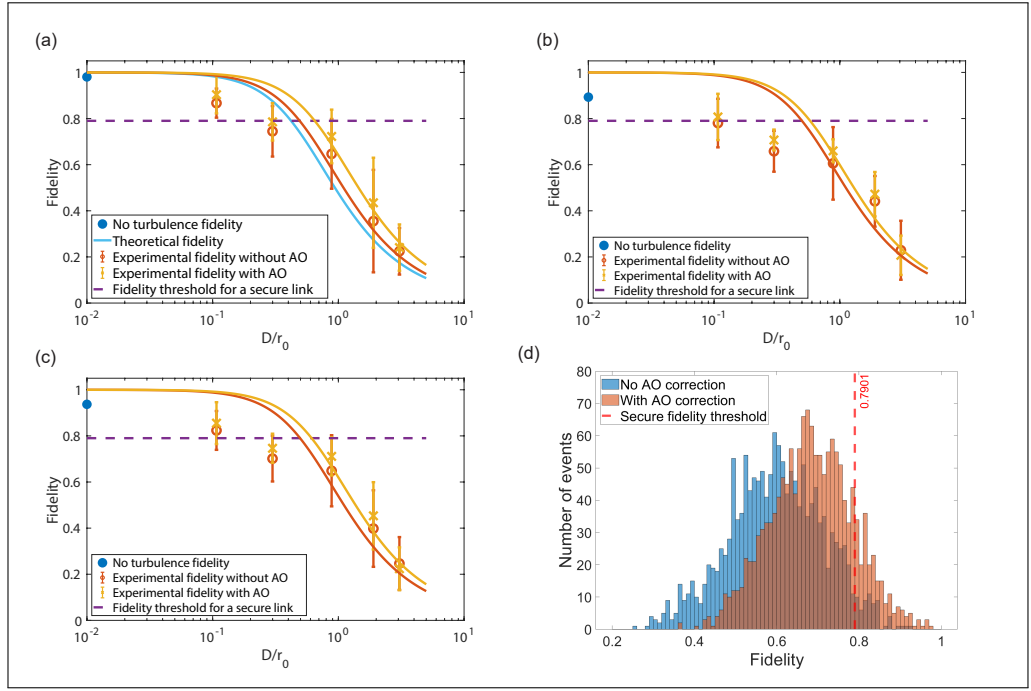}}
\caption{Measured fidelity of lab-scale OAM QKD as functions of turbulence strength and time. (a) measured fidelity of the OAM basis. (b) measured fidelity of the ANG basis. (c) measured fidelity of the MUBs which is the average of (a) and (b). The data point with $D/r_0 = 0.01$ corresponds to the no turbulence case. All the error bars are the measured standard deviation in the fidelity. All the yellow and red curves are least-square fitting results of the measured data using the same model with different coefficient $c$. (d) histogram of the fidelity of the OAM state $\ell$ = 1 with and without AO correction. $D/r_0$ is 0.884. }
\label{fig:fig2}

\end{figure}

The modest improvement in fidelity mainly comes from the limited performance of the AO with an insufficient number of micromirrors on DM and the trade-off between speed and accuracy of the WFS. Our DM only has 32 actuators (6$\times$6 without corners). To avoid being cutoff by the edge, the signal beams are aligned to fall within only the central actuators (3$\times$3 actuators for $\ket{\ell = 1}$) while the beacon beam fills the entire aperture. The inadequate number of actuators can result in a low complexity of the wavefront that can be corrected and a poor accuracy, which means that only the first few orders of Zernike terms can be corrected with limited precision. \par

The trade-off between correction speed and accuracy exists in most AO systems. To provide a fast wavefront measurement, our WFS is set to operate in the high-speed mode. This specific mode can provide a measurement speed up to kHz levels by sacrificing the number of lenslets used to estimate the Zernike coefficients, which indicates that a more accurate wavefront measurement leads to a slower speed of AO compensation. To optimize the performance, one needs to first measure the bandwidth of turbulence to select an AO speed that is fast enough to sample the turbulence. This goal can be achieved by measuring the fidelity fluctuations as a function of time. Usually the atmospheric turbulence fluctuates at tens of hertz, and an AO system should have a correction bandwidth at least equal to the Greenwood frequency, which is around 60 Hz \cite{greenwood1977bandwidth, tyler1994bandwidth}. One also needs to consider the complexity of the Zernike terms that can be corrected by the DM. If the DM has a small number of actuators, one can limit the order of Zernikes measured by the WFS to further improve the speed. Otherwise, the number of Zernike terms measured by WFS should be large enough to avoid the waste of DM correction power. From the number of actuators on the DM, the speed of WFS, and the corresponding number of lenslets, one can determine the beam sizes on the WFS and then use a telescope to relay the field at DM to WFS. Through this procedure, the optimal combination of AO speed and the number of spots on WFS can be found, which will provide the best performance for each specific system. In our case, since the complexity of the DM (6$\times$6 actuators) limits the performance of our AO system, the number of pixels of the WFS is set to a lower level (23$\times$23 lenslets) to provide a faster sampling rate. \par

Another observation is that for a given AO system, the effect of AO compensation varies with the level of turbulence. For our system, the optimal performance occurs when the turbulence is moderate ($D/r_0$ = 0.884). As shown in Fig. 2(d), when the correction is enabled, the average fidelity of the OAM state $\ket{\ell = 1}$ is improved from 64.68$\%$ to 72.24$\%$, and the standard deviation is reduced from 11.24$\%$ to 8.68$\%$ as shown in Fig. 2(d). The probability of events which have an instantaneous fidelity larger than the threshold is increased by a factor of 3.48. In contrast, the improvement in either weak turbulence or strong turbulence is small. This is caused by different reasons in weak turbulence case and strong turbulence case. \par

When the turbulence is weak, the Zernike terms that introduce the majority of the error are tip and tilt, and the high-order terms are usually small and may be negligible compared to the WFS noise. Even though tip and tilt can in principle be easily corrected by two sets of FSM and PD, only one set is involved in our system. Therefore, the propagation direction of the beams leaving the FSM is not under control. After being imaged onto the DM plane, the error in the propagation direction should be corrected by the DM. Apart from tip and tilt, other aberrations in the weak turbulence are not so strong that thus not much error needs to be corrected. This combination of errors seems easy to be corrected. However, due to the inadequate number of actuators on the DM and the noise of the WFS, the AO may introduce some errors to the system that limits the potential fidelity improvement. \par

When strong or deep turbulence (extremely strong turbulence) is present, the high-order Zernike terms contribute more to the errors compared to low-order terms. Not only does the transverse phase profile get disturbed, but also the intensity profile is highly distorted. For example, when strong astigmatism (the 5th and 6th Zernike polynominals in Noll index) is present, the phase singularity at the center of OAM states $\ket{\ell}$ will get fractured into $\ell$ new singularities, and the OAM states will get stretched to elliptical shapes. This phenomenon is observed when $D/r_0$ equals to 1.90 and 3.06, which indicates that the turbulence is strong. In such a case, one simple AO system cannot sufficiently correct the errors in both phase and intensity profiles. Moreover, considering the inadequate number of actuators on the DM, the complexity of the wavefront that the DM can provide is not good enough to correct high-order terms. Therefore, an advanced AO system including multiple conjugate DMs and WFS with fast speed and high resolution is essential to correct strong and deep turbulence, while the exact specifications depends on the level of turbulence \cite{lavery2018vortex}. In contrast, for moderate turbulence, Zernike coefficients are usually large enough so that the WFS can provide a precise measurement, and the DM can also provide a relatively accurate correction. In the meantime, the wavefront complexity is not too great. Considering both effects, our AO correction has adequate performance in the moderate turbulence regime.\par

\subsection{Free-space link across the UR campus}

\begin{figure}[h]

\centering
\fbox{\includegraphics[width=0.95\linewidth]{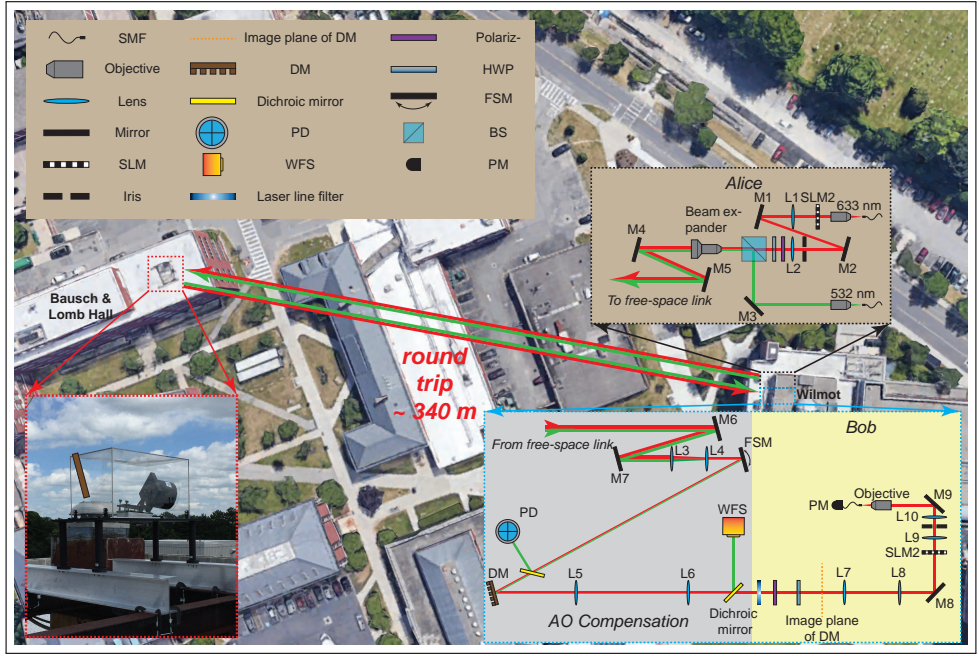}}
\caption{The configuration of a 340 m long cross-campus link. Both Alice and Bob are on the optical table. Since the turbulence in the cross-campus link is not controllable, the turbulence structure number varies between 5.4$\times 10^{-15}$ $\text{m}^{-2/3}$ and 3.2$\times 10^{-14}$ $\text{m}^{-2/3}$.}
\label{fig:fig1}

\end{figure}

We next investigate the performance of AO compensation in a 340 m long free-space link across the University of Rochester (UR). The experimental setup is shown in Fig. 3. The state preparation stage is the same as the setup shown in Fig. 1. After BS, the combined beam is expanded using an achromatic 3$\times$ beam expander (GBE03-A from Thorlabs) and launched to the roof of Bausch $\&$ Lomb Hall through the use of one pair of mirrors (M4 and M5). As shown in the photograph, the hollow retroreflector (\#49-672 from Edmund Optics) and the rotation stage are mounted on an optical breadboard, which is then mounted on the steel and aluminum frames on the roof. To protect the retroreflector, a double protection scheme is used. A high efficiency AR coated protection window is used to seal the front aperture of the retroreflector, which can prevent the formation of dew on the mirrors. The entire system, which is about 35 m above the ground, is also covered by an acrylic protection box (the front side of this box is replaced with a high efficiency window (\#43-975 from Edmund Optics)) to protect the retroreflector and stage from weather. The reflected beams are collected by a pair of mirrors (3 inch clear aperture, M6 and M7), giving a Fresnel number product $N_f$ of the system equal to 4.89. An achromatic lens with 3 inch diameter (L3, $f_3 = 200$ mm) and a negative achromatic lens (L4, $f_4 = -40$ mm) are used to reduce the beam size. After this, the beams are sent to the AO system, which is almost the same as what is shown in Fig. 1 except that the DM has 140 actuators (12$\times$12 without corners, DM140A-35-UM01 from Boston Micromachine). The large DM has more actuators which can provide a better accuracy and complexity in the wavefront correction. To match the size of the clear aperture of DM and WFS, the beam size is reduced by several imaging systems which are not shown in the figure. All the lenses used in the imaging systems are best form spherical singlet lenses to reduce the spherical aberration.\par

\begin{figure}[h!]
\centering
\fbox{\includegraphics[width=0.95\linewidth]{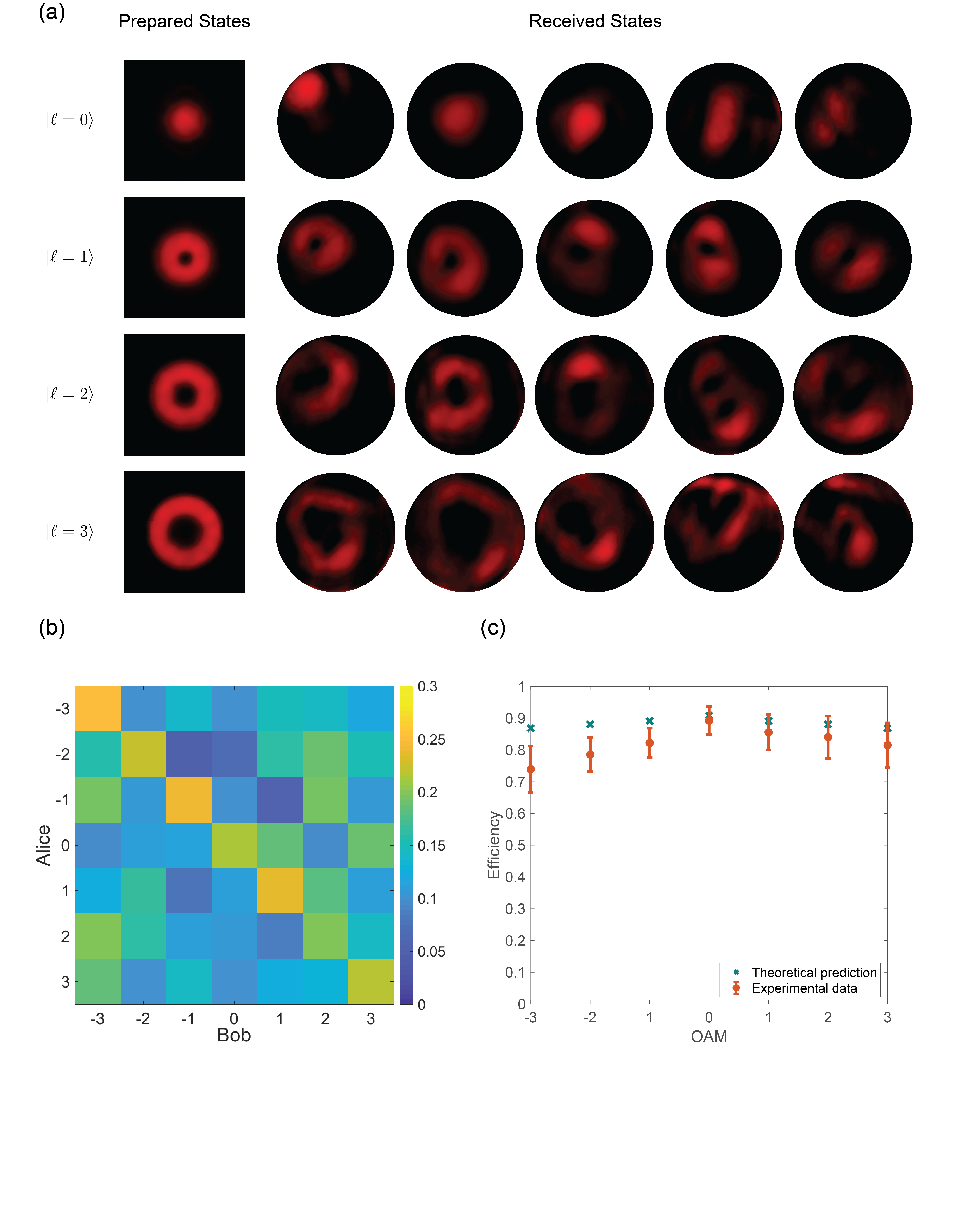}}
\caption{(a) The prepared and received states in the cross-campus link with different OAM values. All the images acquired under same turbulence but at different timepoints during the night showing the distortions on OAM states. Note that the images of the prepared states have been scaled up by a factor of 3. To clearly show the details of the received states, the images of the received states only show the field in the collection aperture, and the intensities of the received $\ket{\ell = 2, 3}$ states are enhanced by a factor of 1.5. (b) crosstalk matrix of the OAM basis after AO correction. The average fidelity is found to be 22.57$\%$. (c) The theoretical and measured mode transmission efficiency. The error bars correspond to the standard deviation in transmission efficiencies. The effect of mode-dependent diffraction has been taken into consideration. All the data shown above are measured in turbulence with $D/r_0 = 2.17$.}
\label{fig:fig1}
\end{figure}

The turbulence structure number $C_n^2$ of our cross-campus link is estimated by calculating the beam wander of the returned $\ket{\ell}$ beam, which yields a $C_n^2$ from 5.4$\times 10^{-15}$ $\text{m}^{-2/3}$ to 3.2$\times 10^{-14}$ $\text{m}^{-2/3}$. This indicates a moderate to strong turbulence \cite{andrews2005laser}. The intensity profiles of the prepared and received states under a turbulence strength $C_n^2 = 1.9\times 10^{-14}$ $\text{m}^{-2/3}$ are shown in Fig. 4(a), and the corresponding $D/r_0$ is 2.17. The intensity profiles are strongly distorted by turbulence so that the original donut shapes of the OAM states are not maintained. One can also see the effects induced by different Zernike terms from these images. For example, the first column of the received states have relatively good intensity profiles but are not at the center of receiver's aperture. These shifts are the result of tip and tilt. The received states in the 4th column show the effects induced by astigmatism. The received states are elongated into elliptical shapes, and the phase singularities are fractured into multiple vortices. In scenarios where multiple Zernike terms dominate the effect, the received states can be highly distorted leading to indistinguishable intensity profiles. For instance, as shown in the 5th column, the $\ket{\ell = 0}$ and $\ket{\ell = 1}$ states are split into 2 separate spots so that the two intensity profiles are similar to each other. \par

The crosstalk matrix and the measured transmission efficiency are shown in Fig. 4(b) and (c). With AO compensation, the fidelity in the OAM basis is only improved from 19.74$\%$ to 22.57$\%$ which is far below the fidelity threshold (76.30$\%$ for $d = 7$ systems), and the improvement in fidelity provided by AO is modest compared to the lab-scale data with a similar turbulence level (for $D/r_0 = 1.90$, the lab-scale fidelity can be improved from 35.53$\%$ to 43.47$\%$). Meanwhile, the transmission efficiency of OAM states, especially the high order terms, also fluctuates significantly. As shown in Fig. 3(c), the theoretical efficiency of the $\ket{\ell = 3}$ state, including the effect of mode-dependent diffraction, should be 86.83$\%$. However, the measured efficiency in the cross-campus link under AO correction is only 73.95$\%$ with 7.32$\%$ standard deviation, which is 12.88$\%$ lower than the expected efficiency. As a comparison, the measured efficiency of a Gaussian state is 89.21$\%$ with little fluctuation, which is only 1.58$\%$ lower than the theoretical efficiency predicted by the simulation. Considering that we are not in the strong or deep turbulence regimes, the low link efficiency with considerable fluctuations and modest improvements with AO are mainly caused by the distorted beacon beam, finite collection aperture size and mode-dependent diffraction \cite{zhao2019performance, tyler2012adaptive}. These effects are only observed in the cross-campus link since the lab-scale link has a sufficiently large Fresnel number product (in our case the $N_f > 270$ in the lab-scale link) and a more stable turbulence strength. The breakdown of the Gaussian state is usually observed in strong and deep turbulence regime but is also occasionally observed in our cross-campus link (the last photograph of the received $\ket{\ell = 0}$ state in Fig. 4(a)), which leads to the failure of precise beam tracking for low Fresnel number product channels. This will then lead to a low link efficiency and a problematic tip-tilt correction, which may introduce additional errors. This effect becomes more severe for a limited Fresnel number product of the link due to finite collection aperture and long propagation distance. In this case, a high-order OAM state arriving at the receiver's aperture will be have a much larger size than the beacon beam due to mode-dependent diffraction. In our case, due to the spherical aberration and defocus, the size of the $\ket{\ell = 3}$ states can vary from 6 cm to more than 7.62 cm, which exceeds the size of collection aperture (effective size is less than 7.62 cm). This indicates that, for a given Fried parameter $r_0$, higher order OAM states with larger cross sections are more distorted than the lower order states. Therefore, a slight mistracking caused by a distorted beacon beam will lead to the cutoff of high-order OAM state and hence a lower efficiency and a much larger error. Meanwhile, due to the mode-dependent diffraction, the size of beacon beam (usually in $\ket{\ell = 0}$ state) cannot match the size of high-order states across the whole link. This indicates that the beacon beam cannot capture all the aberrations that the high-order states experience in the link, which leads to an inefficient and inaccurate AO correction. To solve this, one may need to introduce new protocols to mitigate mode-dependent diffraction, for example as discussed in Ref. \cite{zhao2019performance}. \par

Hence, for free-space OAM QKD channels, a fine beam tracking with a reliable beam stabilization is necessary, which usually requires multiple FSMs and PDs. However, precise beam tracking is still not sufficient to correct the turbulence-induced errors since, in principle, it can only correct lowest Zernike terms (tip and tilt). As we show in the lab-scale link, high-order Zernike terms, which can only be corrected by multiple DMs and WFSs, become dominant under moderate or strong turbulence. That is to say, an advanced AO system should include two basic compositions: at least one set of beam stabilization and tracking system, and one set of advanced wavefront correction system consisting of multiple conjugate DMs and WFSs \cite{GlennBook, GlennSlides, lavery2018vortex}. \par

\section{Discussion} 

To enhance the quality of OAM QKD, one has to mitigate the defects induced by the turbulence and improve the fidelity of the states. One possible solution is increasing the mode spacing in the encoding space. As shown in Figs. 5(a) to (d), under turbulence with $D/r_0 = 1.90$ and no AO correction, we can improve the average fidelity from 35.53$\%$ to 45.72$\%$ by simply encoding information with $\ket{\ell = -4,-2,0,2,4}$ states, which corresponds to a mode spacing of 2. If AO correction is introduced, the average fidelity can be improved from 43.47 to 57.54$\%$. This improvement can be further enhanced if the mode spacing increases. Under the same turbulence, when the mode spacing becomes 4 (i.e. using $\ket{\ell = -4}$, $\ket{\ell = 0}$ and $\ket{\ell = 4}$ states), the fidelity can be improved from 71.49$\%$ to 84.33$\%$ in a $d = 3$ system, which is above the fidelity threshold ($F = 84.05\%$) and a secure channel is achievable (not shown in the figure). However, this solution has two limitations. For a fixed dimension $d$, a large mode spacing involves states with larger $\abs{\ell}$, which will exacerbate the defects induced by mode-dependent diffraction \cite{zhao2019performance} and suffer more turbulence due to the larger beam size. This will result in a lower data rate compared to an encoding system with the same dimensionality but consecutive states. The other limitation is the lack of an efficient sorter to measure the corresponding MUBs. Even though a generic quantum sorter for an arbitrary system has been proposed, implementing such an idea usually requires multiple phase screens, which results in a low overall efficiency and hence a lower key rate and more security loopholes \cite{ionicioiu2016sorting, fontaine2019laguerre}. Therefore, this solution might be not suitable for OAM encoding unless an efficient sorter for OAM MUBs is developed.\par

 \begin{figure}[h!]
\centering
\fbox{\includegraphics[width=0.95\linewidth]{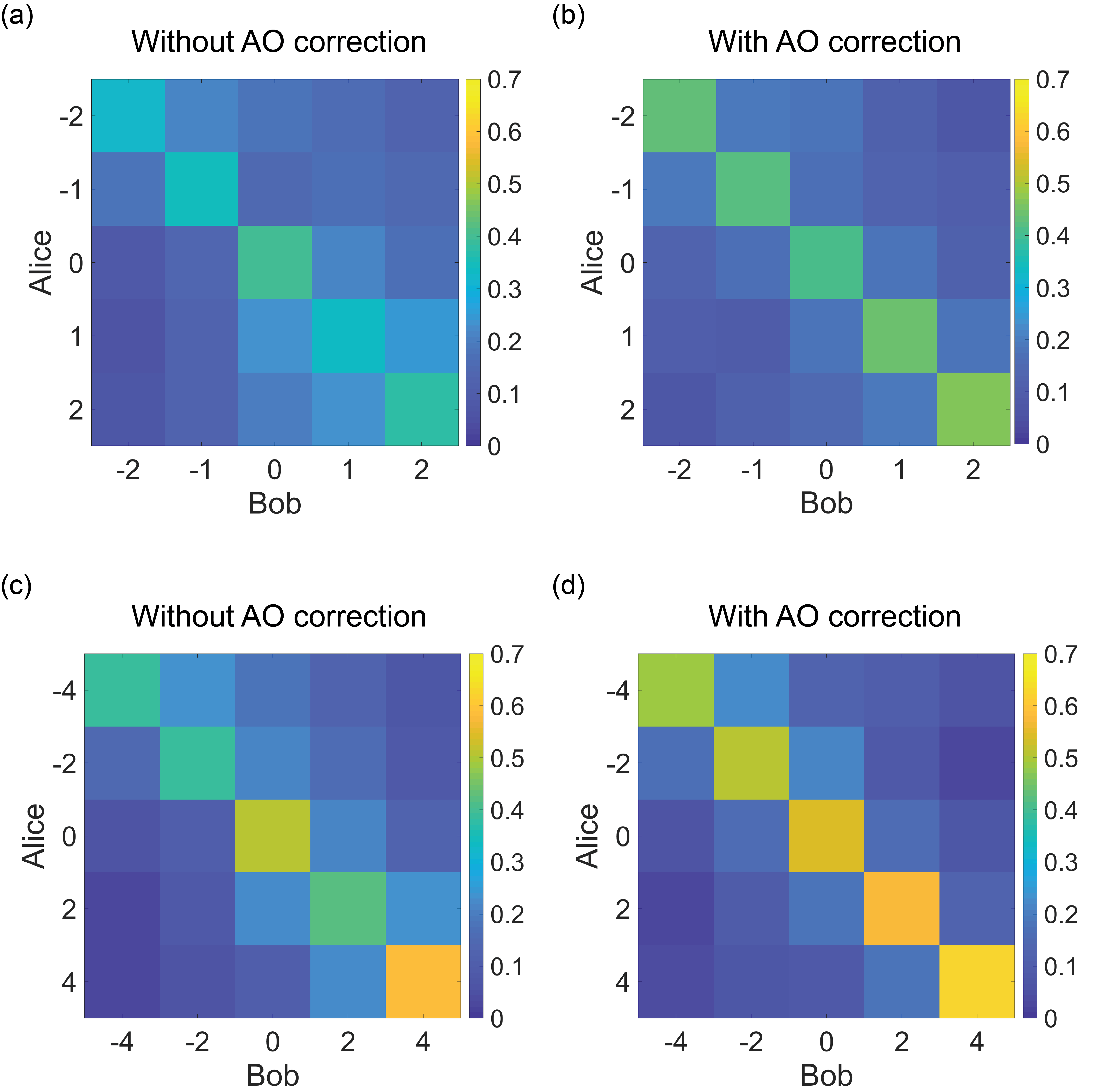}}
\caption{(a) and (b) measured crosstalk matrix of the OAM basis in the lab-scale link without and with AO correction respectively. The average fidelity in (a) is 35.53$\%$ and the fidelity in (b) is 43.47$\%$. The mode spacing is 1 in both cases. (c) and (d) measured crosstalk matrix of OAM basis in the lab-scale link without and with AO correction respectively. The average fidelity in (c) is 45.72$\%$ and the fidelity in (d) is 57.54$\%$. The mode spacing is 2 in both cases. The turbulence level in all figures is $D/r_0 = 1.90$.}
\label{fig:fig1}
\end{figure}

Another approach is to introduce a new degree of freedom, which is robust to turbulence, as the ancillary basis. By doing so, the dimension $d$ can be increased significantly, leading to a more robust encoding system. The most straightforward idea is using polarization as the ancillary basis. The possibility of encoding information on multiple degrees of freedom has been demonstrated in Ref.\cite{zhou2019using}, in which the authors cascade the spatial mode sorter after HWPs and polarizing beam splitters (PBSs) to efficiently measure the received photons. Under turbulence with $D/r_0 = 1.90$, the average fidelity over polarization states is 98.23$\%$. After AO compensation, the average fidelity becomes 98.17$\%$. This indicates that the polarization degree of freedom is robust to turbulence, and will not introduce much additional error to the combined OAM-polarization MUBs. Considering that the error threshold for a $d = 5$ system is 79.01$\%$, the new joint encoding system has a threshold of 73.78$\%$ for $d = 10$, which is 5.23$\%$ lower than before. This improvement will allow two parties to establish a secure channel in our lab-scale link under a turbulence up to $D/r_0 = 0.30$, which would have been impossible without introducing the ancillary basis. The downside of this solution is that the improvement will become relatively small when the original $d$ is large since the error threshold is a logarithmic function of $d$. \par

The last solution is to improve the performance of the entire AO correction system, which is conceptually most straightforward but also most complicated in engineering implementation. As we showed above, a simple AO system consisting of one DM, one WFS, one FSM and one PD will not be sufficient, and the performance of each device should also be improved. Note that the specifications of each device depend on the specifics of the free-space link, and the exact numbers can only be determined when the link parameters are fixed. The DM needs a large number of actuators to provide a high complexity of reconstructed wavefront, which should match the number of lenslets on the WFS and be at least complex enough to correct the dominant Zernike terms. The WFS should have an adequate number of lenslets and a high SNR for an accurate measurement. A fast enough speed, at least faster than the Greenwood frequency (around 60 Hz), to sample the time-varying wavefront and control the compensation loop is also required. The bandwidth of the FSM and PD is usually large enough and the more challenging requirement is the resolution and sensitivity. The FSM should be able to provide a sufficiently small step size but a considerable angular range so that it can accurately direct the beams toward the center of receiver's aperture even when the displacement is large. In a real free-space link, an advanced AO system with multiple devices is necessary to compensate the wavefront error introduced by atmospheric turbulence. The first section of the system should consist of two FSMs, located at the Alice's side, to point the beams at the center of collection aperture. At the Bob's side, the second section of correction system needs two FSMs and two PDs to accurately stabilize the received beams. These two sections jointly provide a precise beam tracking and control, which is essential to the advanced wavefront corrections afterwards \cite{GlennSlides,GlennBook,lavery2018vortex}. An advanced wavefront correction system consisting of multiple conjugate DMs and WFSs with high resolution and large bandwidths is required to mitigate the error in intensity and phase profiles induced by moderate or stronger turbulence. Based on our experimental data and simulation results \cite{GlennBook, GlennSlides}, we believe that an advanced AO correction system as described above is necessary, and in theory it should be able to correct the errors even under strong turbulence. However, the performance of such a system on OAM QKD still needs to be experimentally investigated.\par

Considering the complexity and cost of an advanced AO system as the one described above, it might be difficult to comprehensively build such a system. Therefore, we would suggest the following priority order under a limited budget. Based on our experimental observation, precise beam tracking is usually the first priority of the entire AO system. This conclusion comes from the fact that the OAM states are very sensitive to lateral displacement at the receiver's aperture, and such a tracking system is essential to the rest of the AO system. In addition, a pair of low-noise and high-bandwidth DM and WFS would work better than a pair of high-resolution but noisy low-bandwidth devices. This conclusion comes from the laboratory observation that in weak turbulence cases a pair of low-noise and high-bandwidth DM and WFS can provide better measurement and correction accuracy. However, for strong atmospheric turbulence, both of the pairs will not work and one has to use high-resolution and bandwidth DM and WFS.

\section{Conclusion}
In summary, we study the performance and the possibility of OAM QKD in a turbulent channel under real-time AO correction. The effect of turbulence and the performance of the AO system are quantitatively studied in a lab-scale link under a controllable level of turbulence. We find that, under weak turbulence, real-time AO correction can mitigate the error induced by turbulence and recover the security of the channel. For moderate and strong turbulence, a simple AO system will not be adequate to mitigate the error and an advanced AO system as we described above is required. The performance of an AO system and OAM QKD in a 340 m long free-space link with finite collection aperture is also studied. In additional to the effects we observed in the laboratory measurements, additional errors are induced by the finite collection aperture and mode-dependent diffraction. These errors require one to improve the performance of AO system so that a precise beam tracking and control can be achieved. Finally, we propose three different solutions to improve the performance of OAM QKD over a free-space communication link. We experimentally validate the effectiveness of the first two of the proposed solutions in discussion session, and discuss the possible layout of an advanced AO system. \par

\section{Acknowledgements}
We acknowledge the helpful discussion with Mohammad Mirhosseini, Seyed Mohammad Hashemi Rafsanjani, Omar S. Maga$\tilde{\text{n}}$a-Loaiza and Boshen Gao. J.Z. thanks Myron W. Culver, Tony DiMino, Eric Hebert and Weichen Yao for mounting and aligning the retroreflector. 

\section{Funding}
This work is supported by U.S. Office of Naval Research. R.W.B. acknowledges support from the Canada Excellence Research Chairs Program and the National Science and Engineering Research Council of Canada. B.B. acknowledges the support of the Banting Postdoctoral Fellowship.

\section{Disclosures}
The authors declare no conflicts of interest.

\bibliography{OAM_QKD_TURB}

\begin{thebibliography}{10}
\newcommand{\enquote}[1]{``#1''}

\bibitem{gisin2002quantum}
N.~Gisin, G.~Ribordy, W.~Tittel, and H.~Zbinden, \enquote{Quantum
  cryptography,} {\protect\JournalTitle{Reviews of Modern Physics}}
  \textbf{74}, 145 (2002).

\bibitem{scarani2009security}
V.~Scarani, H.~Bechmann-Pasquinucci, N.~J. Cerf, M.~Du{\v{s}}ek,
  N.~L{\"u}tkenhaus, and M.~Peev, \enquote{The security of practical quantum
  key distribution,} {\protect\JournalTitle{Reviews of Modern Physics}}
  \textbf{81}, 1301 (2009).

\bibitem{lo2014secure}
H.-K. Lo, M.~Curty, and K.~Tamaki, \enquote{Secure quantum key distribution,}
  {\protect\JournalTitle{Nature Photonics}} \textbf{8}, 595 (2014).

\bibitem{bennett1984quantum}
C.~H. Bennett and G.~Brassard, \enquote{Quantum cryptography: public key
  distribution and coin tossing,} {\protect\JournalTitle{Proceedings of the
  IEEE International, Conference on Computers, Systems and Signal Processing}}
  pp. 175--179 (1984).

\bibitem{korzh2015provably}
B.~Korzh, C.~C.~W. Lim, R.~Houlmann, N.~Gisin, M.~J. Li, D.~Nolan,
  B.~Sanguinetti, R.~Thew, and H.~Zbinden, \enquote{Provably secure and
  practical quantum key distribution over 307 km of optical fibre,}
  {\protect\JournalTitle{Nature Photonics}} \textbf{9}, 163 (2015).

\bibitem{yin2016measurement}
H.-L. Yin, T.-Y. Chen, Z.-W. Yu, H.~Liu, L.-X. You, Y.-H. Zhou, S.-J. Chen,
  Y.~Mao, M.-Q. Huang, W.-J. Zhang, H.~Chen, M.~J. Li, D.~Nolan, F.~Zhou,
  X.~Jiang, Z.~Wang, Q.~Zhang, X.-B. Wang, and J.-W. Pan,
  \enquote{Measurement-device-independent quantum key distribution over a 404
  km optical fiber,} {\protect\JournalTitle{Physical review letters}}
  \textbf{117}, 190501 (2016).

\bibitem{liao2017satellite}
S.-K. Liao, W.-Q. Cai, W.-Y. Liu, L.~Zhang, Y.~Li, J.-G. Ren, J.~Yin, Q.~Shen,
  Y.~Cao, Z.-P. Li, F.-Z. Li, X.-W. Chen, L.-H. Sun, J.-J. Jia, J.-C. Wu, X.-J.
  Jiang, J.-F. Wang, Y.-M. Huang, Q.~Wang, Y.-L. Zhou, L.~Deng, T.~Xi, L.~Ma,
  T.~Hu, Q.~Zhang, Y.-A. Chen, N.-L. Liu, X.-B. Wang, Z.-C. Zhu, C.-Y. Lu,
  R.~Shu, C.-Z. Peng, J.-Y. Wang, and J.-W. Pan, \enquote{Satellite-to-ground
  quantum key distribution,} {\protect\JournalTitle{Nature}} \textbf{549}, 43
  (2017).

\bibitem{jin2019genuine}
J.~Jin, J.-P. Bourgoin, R.~Tannous, S.~Agne, C.~J. Pugh, K.~B. Kuntz, B.~L.
  Higgins, and T.~Jennewein, \enquote{Genuine time-bin-encoded quantum key
  distribution over a turbulent depolarizing free-space channel,}
  {\protect\JournalTitle{Optics Express}} \textbf{27}, 37214--37223 (2019).

\bibitem{ji2017towards}
L.~Ji, J.~Gao, A.-L. Yang, Z.~Feng, X.-F. Lin, Z.-G. Li, and X.-M. Jin,
  \enquote{Towards quantum communications in free-space seawater,}
  {\protect\JournalTitle{Optics Express}} \textbf{25}, 19795--19806 (2017).

\bibitem{bouchard2018quantum}
F.~Bouchard, A.~Sit, F.~Hufnagel, A.~Abbas, Y.~Zhang, K.~Heshami, R.~Fickler,
  C.~Marquardt, G.~Leuchs, R.~W. Boyd, and E.~Karimi, \enquote{Quantum
  cryptography with twisted photons through an outdoor underwater channel,}
  {\protect\JournalTitle{Optics Express}} \textbf{26}, 22563--22573 (2018).

\bibitem{ursin2007entanglement}
R.~Ursin, F.~Tiefenbacher, T.~Schmitt-Manderbach, H.~Weier, T.~Scheidl,
  M.~Lindenthal, B.~Blauensteiner, T.~Jennewein, J.~Perdigues, P.~Trojek,
  B.~Ömer, M.~Fürst, M.~Meyenburg, J.~Rarity, Z.~Sodnik, C.~Barbieri,
  H.~Weinfurter, and A.~Zeilinger, \enquote{Entanglement-based quantum
  communication over 144 km,} {\protect\JournalTitle{Nature Physics}}
  \textbf{3}, 481 (2007).

\bibitem{sun2016multiple}
X.~Sun, I.~B. Djordjevic, and M.~A. Neifeld, \enquote{Multiple spatial modes
  based qkd over marine free-space optical channels in the presence of
  atmospheric turbulence,} {\protect\JournalTitle{Optics Express}} \textbf{24},
  27663--27673 (2016).

\bibitem{dousse2010ultrabright}
A.~Dousse, J.~Suffczy{\'n}ski, A.~Beveratos, O.~Krebs, A.~Lema{\^\i}tre,
  I.~Sagnes, J.~Bloch, P.~Voisin, and P.~Senellart, \enquote{Ultrabright source
  of entangled photon pairs,} {\protect\JournalTitle{Nature}} \textbf{466}, 217
  (2010).

\bibitem{steinlechner2012high}
F.~Steinlechner, P.~Trojek, M.~Jofre, H.~Weier, D.~Perez, T.~Jennewein,
  R.~Ursin, J.~Rarity, M.~W. Mitchell, J.~P. Torres, H.~Weinfurter, and
  V.~Pruneri, \enquote{A high-brightness source of polarization-entangled
  photons optimized for applications in free space,}
  {\protect\JournalTitle{Optics Express}} \textbf{20}, 9640--9649 (2012).

\bibitem{cerf2002security}
N.~J. Cerf, M.~Bourennane, A.~Karlsson, and N.~Gisin, \enquote{Security of
  quantum key distribution using d-level systems,}
  {\protect\JournalTitle{Physical Review Letters}} \textbf{88}, 127902 (2002).

\bibitem{sheridan2010security}
L.~Sheridan and V.~Scarani, \enquote{Security proof for quantum key
  distribution using qudit systems,} {\protect\JournalTitle{Physical Review A}}
  \textbf{82}, 030301 (2010).

\bibitem{bradler2016finite}
K.~Br{\'a}dler, M.~Mirhosseini, R.~Fickler, A.~Broadbent, and R.~Boyd,
  \enquote{Finite-key security analysis for multilevel quantum key
  distribution,} {\protect\JournalTitle{New Journal of Physics}} \textbf{18},
  073030 (2016).

\bibitem{vallone2014free}
G.~Vallone, V.~D\text{'}Ambrosio, A.~Sponselli, S.~Slussarenko, L.~Marrucci,
  F.~Sciarrino, and P.~Villoresi, \enquote{Free-space quantum key distribution
  by rotation-invariant twisted photons,} {\protect\JournalTitle{Physical
  Review Letters}} \textbf{113}, 060503 (2014).

\bibitem{mirhosseini2015high}
M.~Mirhosseini, O.~S. Maga{\~n}a-Loaiza, M.~N. O\text{'}Sullivan, B.~Rodenburg,
  M.~Malik, M.~P. Lavery, M.~J. Padgett, D.~J. Gauthier, and R.~W. Boyd,
  \enquote{High-dimensional quantum cryptography with twisted light,}
  {\protect\JournalTitle{New Journal of Physics}} \textbf{17}, 033033 (2015).

\bibitem{sit2017high}
A.~Sit, F.~Bouchard, R.~Fickler, J.~Gagnon-Bischoff, H.~Larocque, K.~Heshami,
  D.~Elser, C.~Peuntinger, K.~G{\"u}nthner, B.~Heim, C.~Marquardt, G.~Leuchs,
  R.~W. Boyd, and E.~Karimi, \enquote{High-dimensional intracity quantum
  cryptography with structured photons,} {\protect\JournalTitle{Optica}}
  \textbf{4}, 1006--1010 (2017).

\bibitem{bouchard2018experimental}
F.~Bouchard, K.~Heshami, D.~England, R.~Fickler, R.~W. Boyd, B.-G. Englert,
  L.~L. S{\'a}nchez-Soto, and E.~Karimi, \enquote{Experimental investigation of
  high-dimensional quantum key distribution protocols with twisted photons,}
  {\protect\JournalTitle{Quantum}} \textbf{2}, 111 (2018).

\bibitem{bouchard2018round}
F.~Bouchard, A.~Sit, K.~Heshami, R.~Fickler, and E.~Karimi,
  \enquote{Round-robin differential-phase-shift quantum key distribution with
  twisted photons,} {\protect\JournalTitle{Physical Review A}} \textbf{98},
  010301 (2018).

\bibitem{wang2012terabit}
J.~Wang, J.-Y. Yang, I.~M. Fazal, N.~Ahmed, Y.~Yan, H.~Huang, Y.~Ren, Y.~Yue,
  S.~Dolinar, M.~Tur, and A.~E. Willner, \enquote{Terabit free-space data
  transmission employing orbital angular momentum multiplexing,}
  {\protect\JournalTitle{Nature Photonics}} \textbf{6}, 488 (2012).

\bibitem{zhu2019compensation}
Z.~Zhu, M.~Janasik, A.~Fyffe, D.~Hay, Y.~Zhou, B.~Kantor, T.~Winder, R.~W.
  Boyd, G.~Leuchs, and Z.~Shi, \enquote{Compensation-free high-capacity
  free-space optical communication using turbulence-resilient vector beams,}
  {\protect\JournalTitle{arXiv preprint arXiv:1910.05406}}  (2019).

\bibitem{allen1992orbital}
L.~Allen, M.~W. Beijersbergen, R.~Spreeuw, and J.~Woerdman, \enquote{Orbital
  angular momentum of light and the transformation of {L}aguerre-{G}aussian
  laser modes,} {\protect\JournalTitle{Physical Review A}} \textbf{45}, 8185
  (1992).

\bibitem{flaes2018robustness}
D.~E.~B. Flaes, J.~Stopka, S.~Turtaev, J.~F. de~Boer, T.~Tyc, and
  T.~{\v{C}}i{\v{z}}m{\'a}r, \enquote{Robustness of light-transport processes
  to bending deformations in graded-index multimode waveguides,}
  {\protect\JournalTitle{Physical Review Letters}} \textbf{120}, 233901 (2018).

\bibitem{cozzolino2019orbital}
D.~Cozzolino, D.~Bacco, B.~Da~Lio, K.~Ingerslev, Y.~Ding, K.~Dalgaard,
  P.~Kristensen, M.~Galili, K.~Rottwitt, S.~Ramachandran, and L.~K.
  Oxenl{\o}we, \enquote{Orbital angular momentum states enabling fiber-based
  high-dimensional quantum communication,} {\protect\JournalTitle{Physical
  Review Applied}} \textbf{11}, 064058 (2019).

\bibitem{tyler2009influence}
G.~A. Tyler and R.~W. Boyd, \enquote{Influence of atmospheric turbulence on the
  propagation of quantum states of light carrying orbital angular momentum,}
  {\protect\JournalTitle{Optics Letters}} \textbf{34}, 142--144 (2009).

\bibitem{tyler2011spatial}
G.~A. Tyler, \enquote{Spatial bandwidth considerations for optical
  communication through a free space propagation link,}
  {\protect\JournalTitle{Optics Letters}} \textbf{36}, 4650--4652 (2011).

\bibitem{ren2013atmospheric}
Y.~Ren, H.~Huang, G.~Xie, N.~Ahmed, Y.~Yan, B.~I. Erkmen, N.~Chandrasekaran,
  M.~P. Lavery, N.~K. Steinhoff, M.~Tur, S.~Dolinar, M.~Neifeld, M.~J. Padgett,
  R.~W. Boyd, J.~H. Shapiro, and A.~E. Willner, \enquote{Atmospheric turbulence
  effects on the performance of a free space optical link employing orbital
  angular momentum multiplexing,} {\protect\JournalTitle{Optics Letters}}
  \textbf{38}, 4062--4065 (2013).

\bibitem{ren2014adaptive}
Y.~Ren, G.~Xie, H.~Huang, N.~Ahmed, Y.~Yan, L.~Li, C.~Bao, M.~P. Lavery,
  M.~Tur, M.~A. Neifeld, R.~W. Boyd, J.~H. Shapiro, and A.~E. Willner,
  \enquote{Adaptive-optics-based simultaneous pre-and post-turbulence
  compensation of multiple orbital-angular-momentum beams in a bidirectional
  free-space optical link,} {\protect\JournalTitle{Optica}} \textbf{1},
  376--382 (2014).

\bibitem{rodenburg2014simulating}
B.~Rodenburg, M.~Mirhosseini, M.~Malik, O.~S. Maga{\~n}a-Loaiza, M.~Yanakas,
  L.~Maher, N.~K. Steinhoff, G.~A. Tyler, and R.~W. Boyd, \enquote{Simulating
  thick atmospheric turbulence in the lab with application to orbital angular
  momentum communication,} {\protect\JournalTitle{New Journal of Physics}}
  \textbf{16}, 033020 (2014).

\bibitem{ren2016experimental}
Y.~Ren, Z.~Wang, P.~Liao, L.~Li, G.~Xie, H.~Huang, Z.~Zhao, Y.~Yan, N.~Ahmed,
  A.~Willner, M.~P.~J. Lavery, N.~Ashrafi, S.~Ashrafi, R.~Bock, M.~Tur, I.~B.
  Djordjevic, M.~A. Neifeld, and A.~E. Willner, \enquote{Experimental
  characterization of a 400 {G}bit/s orbital angular momentum multiplexed
  free-space optical link over 120 m,} {\protect\JournalTitle{Optics Letters}}
  \textbf{41}, 622--625 (2016).

\bibitem{liu2019single}
C.~Liu, K.~Pang, Z.~Zhao, P.~Liao, R.~Zhang, H.~Song, Y.~Cao, J.~Du, L.~Li,
  H.~Song, Y.~Ren, G.~Xie, Y.~Zhao, J.~Zhao, S.~M.~H. Rafsanjani, A.~N.
  Willner, J.~H. Shapiro, R.~W. Boyd, M.~Tur, and A.~E. Willner,
  \enquote{Single-end adaptive optics compensation for emulated turbulence in a
  bi-directional 10-{M}bit/s per channel free-space quantum communication link
  using orbital-angular-momentum encoding,} {\protect\JournalTitle{Research}}
  \textbf{2019}, 8326701 (2019).

\bibitem{zhao2019performance}
J.~Zhao, M.~Mirhosseini, B.~Braverman, Y.~Zhou, S.~M.~H. Rafsanjani, Y.~Ren,
  N.~K. Steinhoff, G.~A. Tyler, A.~E. Willner, and R.~W. Boyd,
  \enquote{Performance analysis of d-dimensional quantum cryptography under
  state-dependent diffraction,} {\protect\JournalTitle{Physical Review A}}
  \textbf{100}, 032319 (2019).

\bibitem{zhao2019experimental}
S.~Zhao, W.~Li, Y.~Shen, Y.~Yu, X.~Han, H.~Zeng, M.~Cai, T.~Qian, S.~Wang,
  Z.~Wang, Y.~Xiao, and Y.~Gu, \enquote{Experimental investigation of quantum
  key distribution over a water channel,} {\protect\JournalTitle{Applied
  Optics}} \textbf{58}, 3902--3907 (2019).

\bibitem{hufnagel2019characterization}
F.~Hufnagel, A.~Sit, F.~Grenapin, F.~Bouchard, K.~Heshami, D.~England,
  Y.~Zhang, B.~J. Sussman, R.~W. Boyd, G.~Leuchs, and E.~Karimi,
  \enquote{Characterization of an underwater channel for quantum communications
  in the {O}ttawa river,} {\protect\JournalTitle{Optics Express}} \textbf{27},
  26346--26354 (2019).

\bibitem{li2019mitigation}
L.~Li, R.~Zhang, P.~Liao, Y.~Cao, H.~Song, Y.~Zhao, J.~Du, Z.~Zhao, C.~Liu,
  K.~Pang, H.~Song, A.~Almaiman, D.~Starodubov, B.~Lynn, R.~Bock, M.~Tur, A.~F.
  Molisch, and A.~E. Willner, \enquote{Mitigation for turbulence effects in a
  40-gbit/s orbital-angular-momentum-multiplexed free-space optical link
  between a ground station and a retro-reflecting uav using mimo equalization,}
  {\protect\JournalTitle{Optics letters}} \textbf{44}, 5181--5184 (2019).

\bibitem{mirhosseini2013efficient}
M.~Mirhosseini, M.~Malik, Z.~Shi, and R.~W. Boyd, \enquote{Efficient separation
  of the orbital angular momentum eigenstates of light,}
  {\protect\JournalTitle{Nature Communications}} \textbf{4}, 2781 (2013).

\bibitem{fu2018realization}
D.~Fu, Y.~Zhou, R.~Qi, S.~Oliver, Y.~Wang, S.~M.~H. Rafsanjani, J.~Zhao,
  M.~Mirhosseini, Z.~Shi, P.~Zhang, and R.~W. Boyd, \enquote{Realization of a
  scalable {L}aguerre--{G}aussian mode sorter based on a robust radial mode
  sorter,} {\protect\JournalTitle{Optics Express}} \textbf{26}, 33057--33065
  (2018).

\bibitem{acin2016necessary}
A.~Ac{\'\i}n, D.~Cavalcanti, E.~Passaro, S.~Pironio, and P.~Skrzypczyk,
  \enquote{Necessary detection efficiencies for secure quantum key distribution
  and bound randomness,} {\protect\JournalTitle{Physical Review A}}
  \textbf{93}, 012319 (2016).

\bibitem{GlennSlides}
G.~A. Tyler, J.~L. Vaughn, and N.~K. Steinhoff, \enquote{High-capacity,
  free-space quantum key distribution based on spatial and polarization
  encoding: Atmospheric considerations interim review,}
  {\protect\JournalTitle{The Optical Science Company}}  (2017).

\bibitem{GlennBook}
G.~A. Tyler, P.~Merritt, R.~Q. Fugate, and T.~J. Brennan, \enquote{Adaptive
  opitcs: Theory and applications,} {\protect\JournalTitle{The Optical Science
  Company}}  (2006).

\bibitem{lavery2018vortex}
M.~P. Lavery, \enquote{Vortex instability in turbulent free-space propagation,}
  {\protect\JournalTitle{New Journal of Physics}} \textbf{20}, 043023 (2018).

\bibitem{ren2017spatially}
Y.~Ren, C.~Liu, K.~Pang, J.~Zhao, Y.~Cao, G.~Xie, L.~Li, P.~Liao, Z.~Zhao,
  M.~Tur, R.~W. Boyd, and A.~E. Willner, \enquote{Spatially multiplexed
  orbital-angular-momentum-encoded single photon and classical channels in a
  free-space optical communication link,} {\protect\JournalTitle{Optics
  Letters}} \textbf{42}, 4881--4884 (2017).

\bibitem{fried1965statistics}
D.~L. Fried, \enquote{Statistics of a geometric representation of wavefront
  distortion,} {\protect\JournalTitle{JOSA}} \textbf{55}, 1427--1435 (1965).

\bibitem{fried1966optical}
D.~L. Fried, \enquote{Optical resolution through a randomly inhomogeneous
  medium for very long and very short exposures,} {\protect\JournalTitle{JOSA}}
  \textbf{56}, 1372--1379 (1966).

\bibitem{andrews2005laser}
L.~C. Andrews and R.~L. Phillips, \emph{Laser beam propagation through random
  media}, vol. 152 (SPIE press Bellingham, WA, 2005).

\bibitem{greenwood1977bandwidth}
D.~P. Greenwood, \enquote{Bandwidth specification for adaptive optics systems,}
  {\protect\JournalTitle{JOSA}} \textbf{67}, 390--393 (1977).

\bibitem{tyler1994bandwidth}
G.~A. Tyler, \enquote{Bandwidth considerations for tracking through
  turbulence,} {\protect\JournalTitle{JOSA A}} \textbf{11}, 358--367 (1994).

\bibitem{tyler2012adaptive}
G.~A. Tyler, \enquote{Adaptive optics compensation for propagation through deep
  turbulence: a study of some interesting approaches,}
  {\protect\JournalTitle{Optical Engineering}} \textbf{52}, 021011 (2012).

\bibitem{ionicioiu2016sorting}
R.~Ionicioiu, \enquote{Sorting quantum systems efficiently,}
  {\protect\JournalTitle{Scientific Reports}} \textbf{6}, 25356 (2016).

\bibitem{fontaine2019laguerre}
N.~K. Fontaine, R.~Ryf, H.~Chen, D.~T. Neilson, K.~Kim, and J.~Carpenter,
  \enquote{Laguerre-{G}aussian mode sorter,} {\protect\JournalTitle{Nature
  communications}} \textbf{10}, 1865 (2019).

\bibitem{zhou2019using}
Y.~Zhou, M.~Mirhosseini, S.~Oliver, J.~Zhao, S.~M.~H. Rafsanjani, M.~P. Lavery,
  A.~E. Willner, and R.~W. Boyd, \enquote{Using all transverse degrees of
  freedom in quantum communications based on a generic mode sorter,}
  {\protect\JournalTitle{Optics Express}} \textbf{27}, 10383--10394 (2019).

\end{thebibliography}


\end{document}